\documentclass[sigconf]{acmart}

\usepackage[frozencache]{minted}
\usepackage{xcolor}
\usepackage{amsmath}
\usepackage{graphicx}
\graphicspath{{figures/}}
\usepackage[pdf]{graphviz}
\usepackage{tikz}
\usepackage{pifont}
\usepackage{subcaption}

\newcommand{\inlinehtml}[1]{\mintinline[fontsize=\small]{html}{#1}}

\AtBeginDocument{%
  \providecommand\BibTeX{{%
    \normalfont B\kern-0.5em{\scshape i\kern-0.25em b}\kern-0.8em\TeX}}}

\copyrightyear{2022}
\acmYear{2022}
\setcopyright{acmlicensed}\acmConference[ITiCSE 2022]{Proceedings of the 27th 
ACM Conference on Innovation and Technology in Computer Science Education Vol 
1}{July 8--13, 2022}{Dublin, Ireland}
\acmBooktitle{Proceedings of the 27th ACM Conference on Innovation and 
Technology in Computer Science Education Vol 1 (ITiCSE 2022), July 8--13, 
2022, 
Dublin, Ireland}
\acmPrice{15.00}
\acmDOI{10.1145/3502718.3524774}
\acmISBN{978-1-4503-9201-3/22/07}

\usepackage{titlesec}
\titlespacing{\section}{2pt}{3pt plus 2pt minus 2pt}{3pt plus 2pt minus 2pt}

\newcommand{\tool}{Proof Blocks}

\newcommand{\pl}{PrairieLearn}

\begin{document}

\title{\tool: Autogradable Scaffolding Activities for Learning to Write 
Proofs}

\author{Seth Poulsen}
\email{sethp3@illinois.edu}
\affiliation{%
  \institution{University of Illinois at Urbana-Champaign}
  \country{USA}
}

\author{Mahesh Viswanathan}
\email{vmahesh@illinois.edu}
\affiliation{%
  \institution{University of Illinois at Urbana-Champaign}
  \country{USA}
}

\author{Geoffrey L. Herman}
\email{glherman@illinois.edu}
\affiliation{%
  \institution{University of Illinois at Urbana-Champaign}
  \country{USA}
}

\author{Matthew West}
\email{mwest@illinois.edu}
\affiliation{%
  \institution{University of Illinois at Urbana-Champaign}
  \country{USA}
}

\renewcommand{\shortauthors}{Poulsen et al.}

\begin{abstract}
In this software tool paper we present \tool, a tool which
enables students to construct mathematical proofs by dragging and dropping
prewritten proof lines into the correct order. We present both implementation
details of the tool, as well as a rich reflection on our experiences using the 
tool in courses with hundreds of students.
\tool{} problems can be graded completely automatically,
enabling students to receive rapid feedback. When writing a problem, the
instructor specifies the dependency graph of the lines of the proof, so that
any correct arrangement of the lines can receive full credit. This innovation
can improve assessment tools by increasing the types of questions
we can ask students about proofs, and potentially give greater access to 
proof knowledge by increasing the amount that students can learn on their own 
with the  help of a computer.
\end{abstract}



\maketitle

\section{Introduction}
Constructing mathematical proofs is one of
the critical, yet difficult skills that students must learn
as a part of the discrete mathematics curriculum.
A panel of 21 experts using a Delphi process agreed that 6 of the 11 most difficult
topics in a typical discrete mathematics course are related to proofs and logic
\cite{goldman2008identifying}.
Proofs and proof techniques are included by the ACM curricular guidelines
as a core knowledge area that should be understood by any student obtaining
a degree in computer engineering, computer science, or software engineering
\cite{acm2013curriculum,acm2014curriculum,acm2016curriculum}.

\begin{figure*}
\centering
\begin{subfigure}[b]{0.62\textwidth}
\begin{tikzpicture}
\node[anchor=north west,inner sep=0] at (0,0) 
{\includegraphics[width=\textwidth]{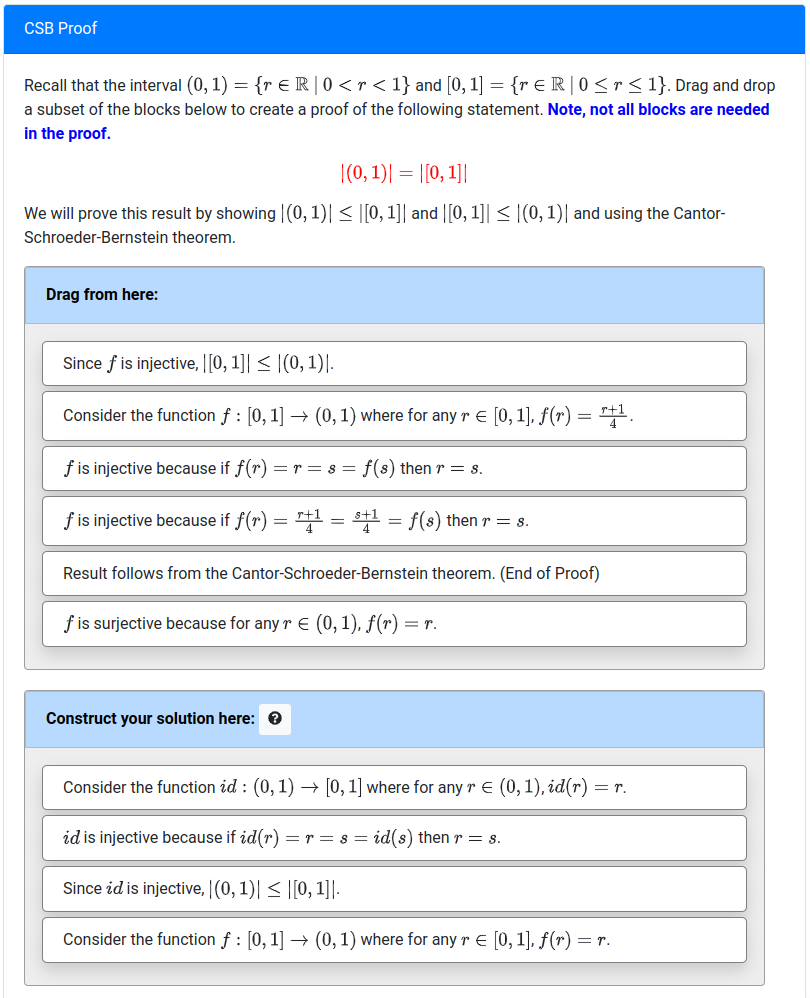}}; 17.3
\node[shape=circle,draw=black,scale=1.2] (1) at (-0.25,-10.75) {1};
\node[shape=circle,draw=black,scale=1.2] (2) at (-0.25,-11.45) {2};
\node[shape=circle,draw=black,scale=1.2] (3) at (-0.25,-12.15) {3};
\node[shape=circle,draw=black,scale=1.2] (4) at (-0.25,-5.67)  {4};
\node[shape=circle,draw=black,scale=1.2] (5) at (-0.25,-7.1){5};
\node[shape=circle,draw=black,scale=1.2] (6) at (-0.25,-4.97)  {6};
\node[shape=circle,draw=black,scale=1.2] (7) at (-0.25,-7.8) {7};
\end{tikzpicture}
\caption{\tool{} screenshot}
\end{subfigure}
\hskip 2em
\begin{subfigure}[b]{0.3\textwidth}
\centering
\raisebox{0.5\height}{
\includegraphics[scale=0.7]{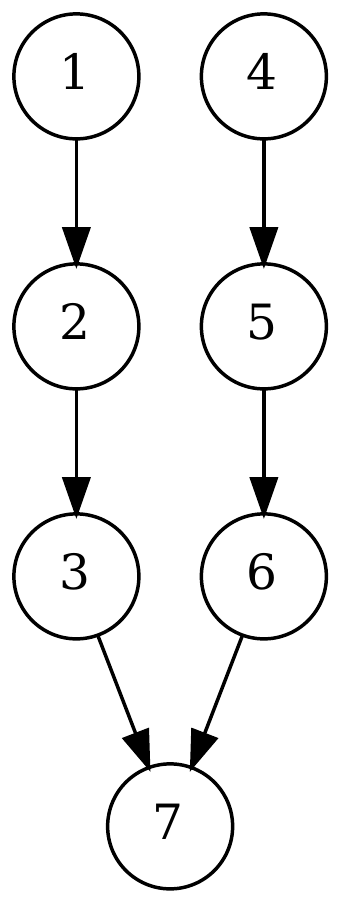}
}
\label{fig:example1-dag}
\caption{Dependency graph of proof lines}
\end{subfigure}
\caption{Example of the \tool~ user interface used by students.
Individual lines of the proof start out shuffled in the starting zone,
and students attempt to drag and drop them into the correct order in 
the target zone. The instructor wrote the
problem with 1, 2, 3, 4, 5, 6, 7 as the intended solution, but the 
\tool~ autograder will also accept any other correct solution as determined by 
the dependency graph 
shown in (b)
For example, both 4, 5, 6, 1, 2, 3, 7 and 1, 4, 2, 3, 5, 6, 7 would also be 
accepted as correct solutions.
}
\label{fig:example1}
\end{figure*}

One problem discrete math instructors face is being able to
provide students with rapid feedback on their
proof writing skills, since proofs must be graded by hand by instructors
or teaching assistants.
With the exception of students who are able to sit down with instructors
during office hours to receive immediate feedback,
most students receive significantly delayed
feedback on the correctness of the proofs which they have constructed
while completing their homework or exams.

What if students were able to receive in-flow automated feedback on their
proofs, just as they are able to with code they write?
Providing students a way to write proofs in such a way that a
computer can give automated feedback  can be a huge advantage. For many students,
this will simply be a convenience factor, but for others, gaining automated
feedback can be a huge step in increasing equity and access in discrete mathematics
education. For example, consider students who are unable to make it to office
hours to receive help due to family commitments, or whose university
courses are understaffed. For these and other populations, automated feedback has
the potential
to make a huge difference by giving them access to feedback they wouldn't have
otherwise received.

Another difficulty for instructors is scaffolding students as they try to make the
 jump
from seeing their instructor write a proof to writing proofs themselves.
To combat this same issue in code writing, researchers have created new
types of learning environments and problems including Parson's Problems
\cite{parsons2006parson} and block programming languages like Scratch
\cite{maloney2010scratch} and Blockly \cite{fraser2015ten}.
The scaffolding provided by both Parson's Problems and block programming
languages have been shown to help students learn more quickly at the
beginning of the learning process \cite{ericson2017solving,weintrop2015block}.
Transitioning from seeing others write proofs to writing them on their own
requires students to use multiple skills, including writing logical
statements and analyzing sequences of logical statements to make sure
that each statement is supported by previous ones.
Due to the complexity of the task, we believe that students should
be given scaffolding for learning to write mathematical proofs, as with
writing code, and they will receive similar benefits.

In this paper, we present \tool, a novel user interface for students to
construct mathematical proofs by dragging and dropping prewritten statements
into the correct order (see Figure~\ref{fig:example1}).

\tool{} allows students to receive instant feedback
on the proofs they have constructed to accelerate the learning process. It also
provides the necessary scaffolding to help students bridge the gap between
seeing others write proofs and writing proofs themselves---reminding students to use
good practices such as
defining variables before using them and being explicit about the proof techniques
being employed.
\tool~ also provide an opportunity for better student assessment, by
providing questions which are, on average, more difficult than multiple choice
questions given to students in a typical discrete mathematics course,
but easier than free response proof writing questions
\cite{poulsen2021evaluating}.

The rest of the paper is organized as follows: we will first discuss related work,
then proceed by explaining the user interface
of \tool~  from both the student and instructor perspective. We will also discuss
our experience using \tool~ in a discrete mathematics course with over 400 students,
and then explain the architecture of the autograder and implications for future work.

The specific contributions of this work are: 
\begin{itemize}
\item A novel grading algorithm for drag-and-drop problems based on a directed 
acyclic graph
\item Application of this algorithm to grading mathematical proofs
\item Insights from experiences using this tool with hundreds of students
\end{itemize}
\begin{figure*}

\includegraphics[width=0.62\textwidth]{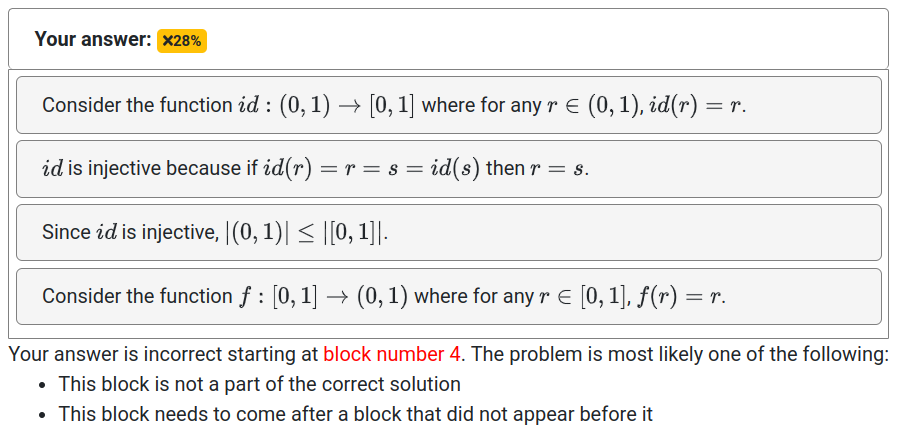}
\caption{Example of feedback given to students working on \tool~ problems.
To avoid giving students so much information that we are not actually
testing their knowledge, they are only told at which line their proof fails,
not the reason why or what the solution is. One area of future research is to
investigate what kind of feedback is best for students to recieve when using
\tool~ as a tool for learning to write proofs.
}
\label{fig:example1-feedback1}
\end{figure*}

\section{Related Work}
Work in intelligent tutors for mathematical proofs goes back to work by John
Anderson and his colleagues on The Geometry Tutor
\cite{anderson1985geometry,anderson1995cognitive,koedinger1990abstract}. More
recently, researchers have created tutors for propositional logic, most notably 
Deep
Thought~\cite{mostafavi2015data,mostafavi2017evolution} and
LogEx~\cite{lodder2020providing-1,lodder2019comparison,lodder2015pilot,loddernodateproving}.
A number of other tools have been created to help students learn to construct
mathematical proofs with the aid of a computer.
Polymorphic Blocks~\cite{lerner2015polymorphic} is a novel user interface
which presents propositions as colorful blocks with uniquely shaped connectors.
The Incredible Proof Machine~\cite{breitner2016visual} guides students through
constructing proofs as graphs.
Jape~\cite{bornat1997jape} is a ``Proof calculator,'' which guides students
through the process of constructing formal proofs in mathematical notation
with the help of the computer, but requires the instructor to implement the 
logics
in a custom language.
MathsTiles~\cite{billingsley2007student} is a block-based programming interface
for constructing proofs for the Isabelle/HOL proof
assistant. Having an open-ended environment where students could
construct arbitrarily complex proofs seems like an advantage, but
user studies showed that students were only successful if they were provided a 
small
instructor-procured subset of blocks.

In reviewing the design of existing tools for computerized proofs it is
clear that there is a tension between two desirable properties: ease of
use for beginners, and ability to handle complex proofs. The tools which
have an elegant, easy to understand interface (Polymorphic Blocks, The Incredible
Proof Machine) only cover formal (and in some cases, simple) logics, limiting their usability for discrete
mathematics courses
where students write informal proofs on a variety of topics from graph theory to number theory.
The tools which can handle an arbitrary complexity of proofs are very
complex and thus difficult and time consuming for students and instructors to use,
especially at the same time as trying to learn to write proofs.
\tool~ solves this problem by allowing
informally written proofs to be automatically graded, making the tool
both easy to use, and able to cover
topics of all level of complexity, including but not limited to
number theory, properties of functions, cardinality, graph theory, Big-O,
and combinatorics.

Anecdotally, we have heard of instructors using scrambled proofs to assess 
student knowledge both in Euclidean geometry and in higher-level mathematics.
In theory, instructors may have offered such questions on paper even before the 
advent of computers, though we can find no explicit record of this.  
Ensley and Winston offer some scrambled proofs in a JavaScript applet as
supplementary material to their discrete mathematics 
textbook~\cite{ensley2005discrete}.
Such questions could also be written using generic drag-and-drop task widgets 
that are present in many learning management systems. However,
the ability use directed acyclic-graph based grading and to demarcate subproofs 
are features unique to \tool~ that we have not seen in any other system, and 
enable assessing proofs which are more complex and use a greater
variety of writing styles.

\section{Development Context}
\tool~ was originally designed for and pilot tested in the discrete mathematics 
course in the computer science department at the University of Illinois at 
Urbana-Champaign during the Fall 2020 semester.
across multiple sections. Most students are freshmen, and take the
course as part of their computer science major, computer science minor, or
computer engineering major. 
The course is designed to prepare students for the theory track in the 
department and usually covers logic, proofs, functions, cardinality, graphs and 
trees, induction, recursion, number theory, probability, basic algorithm 
analysis, and sometimes additional topics as time permits.  
In Fall 2020, the course was held completely online due to the COVID-19 
pandemic. The course was split into 3 sections, each with a different 
instructor,
for a total of over 400 students. The first author of this paper was one of the 
teaching assistants for the course, and the second author was one of the 
instructors.
For more details of the course, and for the results of the 
statistical analyses we performed on data collected from the course, 
see~\cite{poulsen2021evaluating}.

After the semester, we made some small changes based on feedback that we 
recieved, and then the features of \tool{} 
were integrated into the core \pl{} codebase so that they would be open source 
and could be used more 
broadly. \pl{} is in use regularly at about 10 universities, with more 
universities pilot testing. \tool{} is being used by courses in three 
different departments at the University of Illinois, as well as at 
the University of British Columbia and the University of Chicago, and we expect 
these numbers to continue to grow.

\section{User Interface}
\tool~ is built in to \pl{}.
Both the student and teacher user interfaces for creating and
using \tool~ problems are user friendly, and can be used with
almost no training. In an anonymous survey given to our students,
46 out of 51 students responded positively to the statement
``The proof blocks user interface was easy to use,'' with the remaining
5 responding neutrally. Additionally, over two thirds of respondents agreed with
the statements ``Proof Blocks accurately represent my understanding of how to write
proofs,'' and ``Proof Blocks would be a good tool for practicing writing proofs.''
For more detailed survey results, see~\cite{poulsen2021evaluating}.

\subsection{Student Interface}
Figure \ref{fig:example1} shows an example of the \tool~ user
interface seen by students as they work through \tool~ problems.
Individual lines of the proof start out shuffled in the  starting zone,
and students attempt to drag and drop them
into the correct order in the target zone.
Students were able to successfully complete proofs using \tool~
after completing a lecture, worksheet, and homework about proofs,
with no training specifically in how to use the interface.

Figure \ref{fig:example1-feedback1} shows an example of feedback given to students
working on \tool~ problems. This is the feedback that a student would receive
if they were to select ``Save \& Grade'' after having put their \tool~ into the
state shown in Figure \ref{fig:example1}.
To avoid giving students so much information that we are not actually
testing their knowledge, they are only told at which line their proof fails and 
some possible reasons why, not the exact reason why or what the solution is. 
One area of future research is to iterate on what kind of feedback is best for 
students to receive when using \tool~ as a tool for learning to write proofs.

\subsection{Instructor Interface}
In \pl{}, each question written by the instructor will include
(1) an HTML file defining what the students will see, and
(2) a JSON file containing metadata such as the question topic, type, grading
options, and author.
The HTML file may use custom HTML elements defined by \pl{} for writing
homework and exam questions. The HTML is then processed on the backend into
HTML, CSS, and JavaScript before being delivered to the student's browser.

Figure \ref{fig:example1-text} shows the instructor-written HTML code that
generates the \tool~exercise shown in Figure~\ref{fig:example1}. The HTML elements
that are prefixed with ``\inlinehtml{pl}'' have special meaning to \pl{}, which
processes them on the backend before sending the HTML to the client.
The \inlinehtml{pl-question-panel} element notifies \pl{} of the beginning of
a new question. The \inlinehtml{pl-order-blocks} signals to \pl{} to create the
actual
\tool~ user interface, and each \inlinehtml{pl-answer} element inside of it
defines a draggable line of proof.

\begin{figure*}
\centering
\begin{minted}{html}
<pl-question-panel>
  <p>Recall that the interval 
  $(0,1) = \{r \in \mathbb{R}\: |\: 0 \lt r \lt 1\}$ and $[0,1] =  \{r \in \mathbb{R}\: |\: 0 \leq r \leq 1\}$. 
  Drag and drop a subset of the blocks below to create a proof of the following 
  statement. <strong style="color:blue;">Note, not all blocks are needed in the proof.</strong></p>
  <p style="color:red;text-align:center;">$|(0,1)| = |[0,1]|$</p>
  <p>We will prove this result by showing $|(0,1)| \leq |[0,1]|$ and $|[0,1]| \leq |(0,1)|$ and using the 
  Cantor-Schroeder-Bernstein theorem.</p>
</pl-question-panel>

<pl-order-blocks feedback="first-wrong" answers-name="csb-v1" grading-method="dag">
  <pl-answer correct="true" tag="1" depends="">Consider the function $id: (0,1) \to [0,1]$ where for any $r \in (0,1)$, 
    $id(r) = r$.</pl-answer>
  <pl-answer correct="true" tag="2" depends="1">$id$ is injective because if $id(r) = r = s = id(s)$ then $r=s$.</pl-answer>
  <pl-answer correct="true" tag="3" depends="2">Since $id$ is injective, $|(0,1)| \leq |[0,1]|$.</pl-answer>
  <pl-answer correct="true" tag="4" depends="" >Consider the function $f: [0,1] \to (0,1)$ where for any $r \in [0,1]$, 
    $f(r) = \frac{r+1}{4}$.</pl-answer>
  <pl-answer correct="true" tag="5" depends="4">$f$ is injective because if $f(r) = \frac{r+1}{4} = \frac{s+1}{4} = f(s)$ 
    then $r=s$.</pl-answer>
  <pl-answer correct="true" tag="6" depends="5">Since $f$ is injective, $|[0,1]| \leq |(0,1)|$.</pl-answer>
  <pl-answer correct="true" tag="7" depends="3,6">Result follows from the Cantor-Schroeder-Bernstein theorem. 
    (End of Proof)</pl-answer>

  <!-- Distractors -->
  <pl-answer correct="false" tag="" depends="">Consider the function $f: [0,1] \to (0,1)$ where for any $r \in [0,1]$, 
    $f(r) = r$.</pl-answer>
  <pl-answer correct="false" tag="" depends="">$f$ is injective because if $f(r) = r = s = f(s)$ then $r=s$.</pl-answer>
  <pl-answer correct="false" tag="" depends="">$f$ is surjective because for any $r \in (0,1)$, $f(r) = r$.</pl-answer>
</pl-order-blocks>
    
\end{minted}
\caption{The instructor-written HTML code that generates the \tool~exercise
shown in \ref{fig:example1}. The HTML elements that are prefixed with
``\inlinehtml{pl}'' have special meaning to \pl{}, which processes them on the
backend
before sending the HTML to the client. The ``\inlinehtml{depends}'' property on
each ``\inlinehtml{pl-answer}'' element is used to declare the dependency 
between
statements in the proof structure.
}
\label{fig:example1-text}
\end{figure*}

Critically, the instructor writing the problem must specify which lines of
the proof must precede each other line. Though seemingly a small detail,
it is what makes \tool~ such a powerful tool, since it allows instructors
to write proofs with arbitrary English language statements. This overcomes the
proof complexity constraints of earlier student computer proof systems,
and makes it so that students can construct proofs that a computer can
grade at any level of complexity. The proof dependencies are declared using
the ``\inlinehtml{depends}'' attribute. For example, the proof graph for the 
problem
shown in Figure \ref{fig:example1-text} is given in Figure 
\ref{fig:example1}b.

The instructors of the course were able to create new \tool~ questions without
any special training by simply looking at those already created by the authors, only
asking a few questions for clarification about the configuration options,
which could now be answered by looking at the documentation.
An instructor can choose for all of the given lines to be required, or
can add in distractor lines which are not part of the proof.
In our discrete mathematics course, we used test questions both with and without
distractor lines.
Whether or not having distractor lines in the problem leads to better assessment
or learning outcomes is an open question which we leave for future work.

\section{Best Practices For Question Writing}
\label{sec:best-practices}
Our experience using \tool~ with hundreds of students led us to
a few best practices in having \tool~ problems work well for students.

The principal cause for an erroneous {\tool} question is because the instructor
failed to recognize a possible rearrangement of the proof lines that is logically
consistent. This results in a correct student response being incorrectly marked as
faulty by the autograder. Unfortunately, it is easy to make such mistakes when
designing a {\tool} question. These can be avoided if the instructor is aware of the
main reasons this arises, which we outline below. In addition, we recommend that the
instructor ask another member of the course staff who did not design the question,
to solve the problem in different ways \emph{without looking at the source code}. In
our experience, these steps help catch all such mistakes.

One example of this is when the instructor identifies more dependencies
between the proof lines than actually exist. For example an instructor may code up a
problem in a manner which specifies to the autograder that each line in the proof
depends on the line before it. Such strong dependencies are rarely demanded in any
proof. While this is a simple scenario where additional dependencies have been
identified, other cases are more subtle. They often arise because experienced
mathematicians follow \emph{stylistic norms} in addition to logical dependencies
when structuring their proofs. These are so ingrained in a practicing mathematician,
that stylistic norms inadvertently seep in as logical dependencies when coding up a
problem. For example, one often structures proof with subgoals, with the proof of a
new subgoal begun only \emph{after} the proof of the previous subgoal has been
finished. A classical example in a discrete mathematics class is where students are
asked to prove a statement using induction where
the proof of the induction step follows a complete proof of the base case. However,
often there is no logical dependence between the statements in the subproof of each
case. From a logical perspective, the proof statements for each case can be
interleaved in any manner. Of course, emphasizing stylistic norms is just as
important a learning objective, but in that case instructors should be encouraged to
spell this goal out in the problem statement. To avoid such mistakes, after coding a
{\tool} question, we encourage instructors to examine the dependencies of each line
in the coded problem in isolation, without the large proof context.

Another common cause for errors arises in proofs that contain many algebraic
manipulation steps. In informal proof writing, it is often acceptable to skip
intermediate steps of algebraic manipulation. Coding a question in a manner that
demands all the steps leads to student complaints about the autograder. There are
two ways to address this problem. One is to write multiple algebraic simplification
steps in a single proof statement in the problem. The second, and probably the best,
is to avoid having any distractors in the problem, and notify the student that
\emph{all} blocks should be used to construct a correct proof.

The last cause for an error could be distractors. When designing a question, it is
useful to remember that none of the distractors should be part of \emph{any} correct
proof. A common mistake is to have distractors that are superfluous to the
correct proof; this is a problem because we can write logically correct proofs that
have additional statements that do not contribute to the end goal. Thus, it is
important to ensure that adding any distractor would result in a logically
inconsistent argument. One simple way to ensure this is to have each distractor 
(on
its own) be a logically inconsistent statement. Even though this might seem 
like an
easy distractor for a student to avoid, in practice we have found that students are
nonetheless confounded by such distractors.
In the future, we hope to extend the \tool{} grader so that it can also handle 
having lines which can optionally be part of a correct proof.

\section{Autograder}
The autograder is currently built in to \pl{}, but the core algorithm
is about 100 
lines of Python code that could be made to work
with an alternative frontend, or reimplemented in any other
language.

While creating the tool, we recognized that it would be a poor student
experience if the student was expected to place the lines of the proof
in the exact order which the instructor first wrote them,
because in many mathematical proofs, certain lines can
be permuted without affecting the correctness of the proof.
It would also be a poor user experience for the instructor if
they had to explicitly declare every possible correct answer to each
question. This led us to our current grading
scheme, which is based on the dependency graph of the lines in
the proof, which is a directed acyclic graph (DAG).
The instructor simply declares the
dependency graph of statements in the proof, and then the grader
will accept any correct permutation of the lines.

In the basic case, where a proof
has no subproofs like the example in Figure \ref{fig:example1},
checking if a proof is correct is equivalent to checking if the
student ordering of the lines is a topological sort of the DAG.
A more rigorous treatment of the grading algorithm, and the details of our 
edit-distance based partial 
credit algorithm can be seen in~\cite{poulsen2022efficient}.

\subsection{Subproofs}
Even in an introductory discrete mathematics course, an instructor may
want to use proofs that have cases. For example, using cases
to prove an ``or'' statement, or proof by induction.
Here each subproof is a connected subgraph of the entire proof graph.
In such cases,
checking for topological sorting of the proof DAG is insufficient,
because this would allow for intermixing of statements from separate
subproofs in a nonsensical fashion.
A correct proof is a topological sort of the lines of the
proof with the added condition that the lines of each subproof must
be listed contiguously.
Therefore, there is an extra check which ensures
that once a given subproof is started, it is finished before any lines
from outside the subproof appear.

To write a question with a subproof, the instructor wraps each subproof in a 
\inlinehtml{pl-block-group} element. The \inlinehtml{pl-block-group} element 
may then be given its own tag, so that lines that logically depend on the 
entire subproof can refer to them in their \inlinehtml{depends} attribute.
For more details and examples of problems with subproofs, see the
\tool~ documentation~\cite{orderblocksdocs}.
As noted in Section
\ref{sec:best-practices}, it is important to note that subproofs declared only 
for stylistic, and not logical, reasons can be misleading for student unless 
they are explicitly notified of the style which they are to follow.

\section{Evaluation}
Using data from hundreds of student exams from the discrete math course in Fall
2020, we have shown that \tool~ problems provide about as much information about
student knowledge as written proof
problems do. We have also shown that as test questions, \tool~ problems are in fact
easier than written proofs, which are often very difficult. An anonymous
survey given to these students showed that students felt that \tool~ problems
accurately represented their ability to write proofs, and that the user interface
was easy to use. Full details
of this evaluation can be seen in \cite{poulsen2021evaluating}. Ongoing 
evaluation work seeks to explore the possibility that Proof Blocks can help 
students learn more efficiently than writing proofs from scratch, just as 
Parson's Problems can help
students learn more efficiently than writing code from scratch
~\cite{ericson2017solving}.

\section{Adopting \tool}
To use \tool~ with your students, start by following the onboarding
instructions for \pl{}~\cite{pldocs}.
Once familiar with the basic workings of \pl{},
follow the documentation for writing \tool~ questions~\cite{orderblocksdocs}.
More example problems can be found in the documentation and example courses.
\pl{} is in the process of integrating with Learning Tools Interoperability
\cite{severance2010ims} to enable easier sharing of student data across learning
platforms.
Feel free to reach out to the authors with any questions, or about the possibility
of adding \tool~ support on other platforms.

\section{Limitations}
The key limitation of \tool~ is that it restricts what students can do, only allowing
them to place prewritten lines into their proof rather than allowing them to write
whatever they want. As with Parson's Problems and
block based programming languages, we expect that there is a certain skill level at
which \tool~ will become a hindrance rather than a help to students, but this is of
course expected for all forms of education scaffolding. Similarly, we believe that
\tool~ will can be a huge help for students who are just getting started in learning
to write proofs.
Another limitation is that proofs will be graded correctly only as long as the
instructor correctly codes the question---but this is really no worse than most other types of exam questions
given to students. Finally, \tool~ is currently only usable within the \pl{}.
Ongoing efforts to improve interoperability between \pl{} and other learning
platforms will help ease adoption.

\section{Future Work and implications}
The versatility of the \tool~ platform makes it ideally suited for
many future avenues of research.
Next, we would like to enable automatic generation of \tool~
problems so that students can have essentially unlimited practice.
We will also want to research a way
to predict the difficulty of a given generated problem, so students can be
guided through questions of varying difficulty as the learn, and for fairness
on assessments. Beyond discrete mathematics, there are great
possibilities in using \tool{} problems for other courses involving proofs
such as algorithms courses, or even high school geometry.


As noted, \tool~ are a good way to bridge the gap between students
learning to read and write proofs. To give further support to students
as they learn to write proofs, we can try variations on \tool. For example,
we could have students drag and drop lines of a proof which are mostly prewritten,
but have some blanks for the students to fill in,
much as Weinman et al. have done with their ``Faded Parson's Problems''
\cite{weinman2020exploring}

\tool~ can increase access to proof knowledge
by helping students gain more rapid feedback on the proofs they write. Additionaly,
it can improve assessment tools by increasing the types and difficulty levels of
questions we can ask students about proofs, and save many hours of instructor grading
time which can be reallocated to office hours or other effective means of helping
students
\cite{poulsen2021evaluating}.
There has been some evidence that mathematics is acting as a gatekeeper to learning
programming, and that it doesn't actually predict performance in software developers
\cite{ensmenger2012computer}.
Furthermore, many people going into software development study curricula that
involve less math than a standard computer science curricula.
\tool~ can also provide
a solution in this case: rather than teaching less mathematics, \tool~
provides a middle ground. Students can be introduced to logical thinking and proof
writing in a gentler way, potentially reducing the gatekeeping of mathematics while
helping students learn the content.

\begin{acks}
We would like to thank Benjamin Cosman, Patrick Lin, and Yael Gertner for being 
willing to test early versions of \tool{} with their students, and the 
Computers \& Education research group at the University of Illinois for 
feedback on earlier versions of this paper.
\end{acks}

\bibliographystyle{ACM-Reference-Format}
\bibliography{references}


\begin{thebibliography}{30}


\ifx \showCODEN    \undefined \def \showCODEN     #1{\unskip}     \fi
\ifx \showDOI      \undefined \def \showDOI       #1{#1}\fi
\ifx \showISBNx    \undefined \def \showISBNx     #1{\unskip}     \fi
\ifx \showISBNxiii \undefined \def \showISBNxiii  #1{\unskip}     \fi
\ifx \showISSN     \undefined \def \showISSN      #1{\unskip}     \fi
\ifx \showLCCN     \undefined \def \showLCCN      #1{\unskip}     \fi
\ifx \shownote     \undefined \def \shownote      #1{#1}          \fi
\ifx \showarticletitle \undefined \def \showarticletitle #1{#1}   \fi
\ifx \showURL      \undefined \def \showURL       {\relax}        \fi
\providecommand\bibfield[2]{#2}
\providecommand\bibinfo[2]{#2}
\providecommand\natexlab[1]{#1}
\providecommand\showeprint[2][]{arXiv:#2}

\bibitem[\protect\citeauthoryear{Anderson, Boyle, and Yost}{Anderson
  et~al\mbox{.}}{1985}]%
        {anderson1985geometry}
\bibfield{author}{\bibinfo{person}{JR Anderson}, \bibinfo{person}{CF Boyle},
  {and} \bibinfo{person}{G Yost}.} \bibinfo{year}{1985}\natexlab{}.
\newblock \bibinfo{title}{The Geometry Tutor, proc. of 9th Internation Joint
  Conference on Artificial Intelligence}.
\newblock
\newblock


\bibitem[\protect\citeauthoryear{Anderson, Corbett, Koedinger, and
  Pelletier}{Anderson et~al\mbox{.}}{1995}]%
        {anderson1995cognitive}
\bibfield{author}{\bibinfo{person}{John~R Anderson}, \bibinfo{person}{Albert~T
  Corbett}, \bibinfo{person}{Kenneth~R Koedinger}, {and} \bibinfo{person}{Ray
  Pelletier}.} \bibinfo{year}{1995}\natexlab{}.
\newblock \showarticletitle{Cognitive tutors: Lessons learned}.
\newblock \bibinfo{journal}{\emph{The journal of the learning sciences}}
  \bibinfo{volume}{4}, \bibinfo{number}{2} (\bibinfo{year}{1995}),
  \bibinfo{pages}{167--207}.
\newblock


\bibitem[\protect\citeauthoryear{Billingsley and Robinson}{Billingsley and
  Robinson}{2007}]%
        {billingsley2007student}
\bibfield{author}{\bibinfo{person}{William Billingsley} {and}
  \bibinfo{person}{Peter Robinson}.} \bibinfo{year}{2007}\natexlab{}.
\newblock \showarticletitle{Student proof exercises using MathsTiles and
  Isabelle/HOL in an intelligent book}.
\newblock \bibinfo{journal}{\emph{Journal of Automated Reasoning}}
  \bibinfo{volume}{39}, \bibinfo{number}{2} (\bibinfo{year}{2007}),
  \bibinfo{pages}{181--218}.
\newblock


\bibitem[\protect\citeauthoryear{Bornat and Sufrin}{Bornat and Sufrin}{1997}]%
        {bornat1997jape}
\bibfield{author}{\bibinfo{person}{Richard Bornat} {and}
  \bibinfo{person}{Bernard Sufrin}.} \bibinfo{year}{1997}\natexlab{}.
\newblock \showarticletitle{Jape: A calculator for animating proof-on-paper}.
  In \bibinfo{booktitle}{\emph{International Conference on Automated
  Deduction}}. Springer, \bibinfo{pages}{412--415}.
\newblock


\bibitem[\protect\citeauthoryear{Breitner}{Breitner}{2016}]%
        {breitner2016visual}
\bibfield{author}{\bibinfo{person}{Joachim Breitner}.}
  \bibinfo{year}{2016}\natexlab{}.
\newblock \showarticletitle{Visual theorem proving with the Incredible Proof
  Machine}. In \bibinfo{booktitle}{\emph{International Conference on
  Interactive Theorem Proving}}. Springer, \bibinfo{pages}{123--139}.
\newblock


\bibitem[\protect\citeauthoryear{Ensley and Crawley}{Ensley and
  Crawley}{2005}]%
        {ensley2005discrete}
\bibfield{author}{\bibinfo{person}{Douglas~E Ensley} {and}
  \bibinfo{person}{J~Winston Crawley}.} \bibinfo{year}{2005}\natexlab{}.
\newblock \bibinfo{booktitle}{\emph{Discrete mathematics: mathematical
  reasoning and proof with puzzles, patterns, and games}}.
\newblock \bibinfo{publisher}{John Wiley \& Sons}.
\newblock


\bibitem[\protect\citeauthoryear{Ensmenger}{Ensmenger}{2012}]%
        {ensmenger2012computer}
\bibfield{author}{\bibinfo{person}{Nathan~L Ensmenger}.}
  \bibinfo{year}{2012}\natexlab{}.
\newblock \bibinfo{booktitle}{\emph{The computer boys take over: Computers,
  programmers, and the politics of technical expertise}}.
\newblock \bibinfo{publisher}{Mit Press}.
\newblock


\bibitem[\protect\citeauthoryear{Ericson, Margulieux, and Rick}{Ericson
  et~al\mbox{.}}{2017}]%
        {ericson2017solving}
\bibfield{author}{\bibinfo{person}{Barbara~J Ericson},
  \bibinfo{person}{Lauren~E Margulieux}, {and} \bibinfo{person}{Jochen Rick}.}
  \bibinfo{year}{2017}\natexlab{}.
\newblock \showarticletitle{Solving parsons problems versus fixing and writing
  code}. In \bibinfo{booktitle}{\emph{Proceedings of the 17th Koli Calling
  International Conference on Computing Education Research}}.
  \bibinfo{pages}{20--29}.
\newblock


\bibitem[\protect\citeauthoryear{{Fraser}}{{Fraser}}{2015}]%
        {fraser2015ten}
\bibfield{author}{\bibinfo{person}{N. {Fraser}}.}
  \bibinfo{year}{2015}\natexlab{}.
\newblock \showarticletitle{Ten things we've learned from Blockly}. In
  \bibinfo{booktitle}{\emph{2015 IEEE Blocks and Beyond Workshop (Blocks and
  Beyond)}}. \bibinfo{pages}{49--50}.
\newblock
\urldef\tempurl%
\url{https://doi.org/10.1109/BLOCKS.2015.7369000}
\showDOI{\tempurl}


\bibitem[\protect\citeauthoryear{Goldman, Gross, Heeren, Herman, Kaczmarczyk,
  Loui, and Zilles}{Goldman et~al\mbox{.}}{2008}]%
        {goldman2008identifying}
\bibfield{author}{\bibinfo{person}{Ken Goldman}, \bibinfo{person}{Paul Gross},
  \bibinfo{person}{Cinda Heeren}, \bibinfo{person}{Geoffrey Herman},
  \bibinfo{person}{Lisa Kaczmarczyk}, \bibinfo{person}{Michael~C Loui}, {and}
  \bibinfo{person}{Craig Zilles}.} \bibinfo{year}{2008}\natexlab{}.
\newblock \showarticletitle{Identifying important and difficult concepts in
  introductory computing courses using a delphi process}. In
  \bibinfo{booktitle}{\emph{Proceedings of the 39th SIGCSE technical symposium
  on Computer science education}}. \bibinfo{pages}{256--260}.
\newblock


\bibitem[\protect\citeauthoryear{Joint Task Force~on Computing~Curricula and
  Society}{Joint Task Force~on Computing~Curricula and Society}{2013}]%
        {acm2013curriculum}
\bibfield{author}{\bibinfo{person}{Association for Computing Machinery~(ACM)
  Joint Task Force~on Computing~Curricula} {and} \bibinfo{person}{IEEE~Computer
  Society}.} \bibinfo{year}{2013}\natexlab{}.
\newblock \bibinfo{booktitle}{\emph{Computer Science Curricula 2013: Curriculum
  Guidelines for Undergraduate Degree Programs in Computer Science}}.
\newblock \bibinfo{publisher}{Association for Computing Machinery},
  \bibinfo{address}{New York, NY, USA}.
\newblock
\showISBNx{9781450323093}


\bibitem[\protect\citeauthoryear{Koedinger and Anderson}{Koedinger and
  Anderson}{1990}]%
        {koedinger1990abstract}
\bibfield{author}{\bibinfo{person}{Kenneth~R Koedinger} {and}
  \bibinfo{person}{John~R Anderson}.} \bibinfo{year}{1990}\natexlab{}.
\newblock \showarticletitle{Abstract planning and perceptual chunks: Elements
  of expertise in geometry}.
\newblock \bibinfo{journal}{\emph{Cognitive Science}} \bibinfo{volume}{14},
  \bibinfo{number}{4} (\bibinfo{year}{1990}), \bibinfo{pages}{511--550}.
\newblock


\bibitem[\protect\citeauthoryear{Lerner, Foster, and Griswold}{Lerner
  et~al\mbox{.}}{2015}]%
        {lerner2015polymorphic}
\bibfield{author}{\bibinfo{person}{Sorin Lerner}, \bibinfo{person}{Stephen~R
  Foster}, {and} \bibinfo{person}{William~G Griswold}.}
  \bibinfo{year}{2015}\natexlab{}.
\newblock \showarticletitle{Polymorphic blocks: Formalism-inspired UI for
  structured connectors}. In \bibinfo{booktitle}{\emph{Proceedings of the 33rd
  Annual ACM Conference on Human Factors in Computing Systems}}.
  \bibinfo{pages}{3063--3072}.
\newblock


\bibitem[\protect\citeauthoryear{Lodder, Heeren, and Jeuring}{Lodder
  et~al\mbox{.}}{[n.d.]}]%
        {loddernodateproving}
\bibfield{author}{\bibinfo{person}{Josje Lodder}, \bibinfo{person}{Bastiaan
  Heeren}, {and} \bibinfo{person}{Johan Jeuring}.}
  \bibinfo{year}{[n.d.]}\natexlab{}.
\newblock \showarticletitle{Proving the equivalence of two logical formulae
  with {LogEx}}.
\newblock  (\bibinfo{year}{[n.\,d.]}), \bibinfo{pages}{3}.
\newblock


\bibitem[\protect\citeauthoryear{Lodder, Heeren, and Jeuring}{Lodder
  et~al\mbox{.}}{2015}]%
        {lodder2015pilot}
\bibfield{author}{\bibinfo{person}{Josje Lodder}, \bibinfo{person}{Bastiaan
  Heeren}, {and} \bibinfo{person}{Johan Jeuring}.}
  \bibinfo{year}{2015}\natexlab{}.
\newblock \showarticletitle{A pilot study of the use of {LogEx}, lessons
  learned}.
\newblock \bibinfo{journal}{\emph{arXiv:1507.03671 [cs]}} (\bibinfo{date}{July}
  \bibinfo{year}{2015}).
\newblock
\urldef\tempurl%
\url{http://arxiv.org/abs/1507.03671}
\showURL{%
\tempurl}
\newblock
\shownote{arXiv: 1507.03671.}


\bibitem[\protect\citeauthoryear{Lodder, Heeren, and Jeuring}{Lodder
  et~al\mbox{.}}{2019}]%
        {lodder2019comparison}
\bibfield{author}{\bibinfo{person}{Josje Lodder}, \bibinfo{person}{Bastiaan
  Heeren}, {and} \bibinfo{person}{Johan Jeuring}.}
  \bibinfo{year}{2019}\natexlab{}.
\newblock \showarticletitle{A comparison of elaborated and restricted feedback
  in {LogEx}, a tool for teaching rewriting logical formulae}.
\newblock \bibinfo{journal}{\emph{Journal of Computer Assisted Learning}}
  \bibinfo{volume}{35}, \bibinfo{number}{5} (\bibinfo{year}{2019}),
  \bibinfo{pages}{620--632}.
\newblock
\showISSN{1365-2729}
\urldef\tempurl%
\url{https://doi.org/10.1111/jcal.12365}
\showDOI{\tempurl}
\newblock
\shownote{\_eprint:
  https://onlinelibrary.wiley.com/doi/pdf/10.1111/jcal.12365.}


\bibitem[\protect\citeauthoryear{Lodder, Heeren, and Jeuring}{Lodder
  et~al\mbox{.}}{2020}]%
        {lodder2020providing-1}
\bibfield{author}{\bibinfo{person}{Josje Lodder}, \bibinfo{person}{Bastiaan
  Heeren}, {and} \bibinfo{person}{Johan Jeuring}.}
  \bibinfo{year}{2020}\natexlab{}.
\newblock \showarticletitle{Providing {Hints}, {Next} {Steps} and {Feedback} in
  a {Tutoring} {System} for {Structural} {Induction}}.
\newblock \bibinfo{journal}{\emph{Electronic Proceedings in Theoretical
  Computer Science}}  \bibinfo{volume}{313} (\bibinfo{date}{Feb.}
  \bibinfo{year}{2020}), \bibinfo{pages}{17--34}.
\newblock
\showISSN{2075-2180}
\urldef\tempurl%
\url{https://doi.org/10.4204/EPTCS.313.2}
\showDOI{\tempurl}
\newblock
\shownote{arXiv: 2002.12552.}


\bibitem[\protect\citeauthoryear{Maloney, Resnick, Rusk, Silverman, and
  Eastmond}{Maloney et~al\mbox{.}}{2010}]%
        {maloney2010scratch}
\bibfield{author}{\bibinfo{person}{John Maloney}, \bibinfo{person}{Mitchel
  Resnick}, \bibinfo{person}{Natalie Rusk}, \bibinfo{person}{Brian Silverman},
  {and} \bibinfo{person}{Evelyn Eastmond}.} \bibinfo{year}{2010}\natexlab{}.
\newblock \showarticletitle{The scratch programming language and environment}.
\newblock \bibinfo{journal}{\emph{ACM Transactions on Computing Education
  (TOCE)}} \bibinfo{volume}{10}, \bibinfo{number}{4} (\bibinfo{year}{2010}),
  \bibinfo{pages}{1--15}.
\newblock


\bibitem[\protect\citeauthoryear{Mostafavi and Barnes}{Mostafavi and
  Barnes}{2017}]%
        {mostafavi2017evolution}
\bibfield{author}{\bibinfo{person}{Behrooz Mostafavi} {and}
  \bibinfo{person}{Tiffany Barnes}.} \bibinfo{year}{2017}\natexlab{}.
\newblock \showarticletitle{Evolution of an {Intelligent} {Deductive} {Logic}
  {Tutor} {Using} {Data}-{Driven} {Elements}}.
\newblock \bibinfo{journal}{\emph{International Journal of Artificial
  Intelligence in Education}} \bibinfo{volume}{27}, \bibinfo{number}{1}
  (\bibinfo{date}{March} \bibinfo{year}{2017}), \bibinfo{pages}{5--36}.
\newblock
\showISSN{1560-4292, 1560-4306}
\urldef\tempurl%
\url{https://doi.org/10.1007/s40593-016-0112-1}
\showDOI{\tempurl}


\bibitem[\protect\citeauthoryear{Mostafavi, Zhou, Lynch, Chi, and
  Barnes}{Mostafavi et~al\mbox{.}}{2015}]%
        {mostafavi2015data}
\bibfield{author}{\bibinfo{person}{Behrooz Mostafavi}, \bibinfo{person}{Guojing
  Zhou}, \bibinfo{person}{Collin Lynch}, \bibinfo{person}{Min Chi}, {and}
  \bibinfo{person}{Tiffany Barnes}.} \bibinfo{year}{2015}\natexlab{}.
\newblock \showarticletitle{Data-{Driven} {Worked} {Examples} {Improve}
  {Retention} and {Completion} in a {Logic} {Tutor}}.
\newblock In \bibinfo{booktitle}{\emph{Artificial {Intelligence} in
  {Education}}}, \bibfield{editor}{\bibinfo{person}{Cristina Conati},
  \bibinfo{person}{Neil Heffernan}, \bibinfo{person}{Antonija Mitrovic}, {and}
  \bibinfo{person}{M.~Felisa Verdejo}} (Eds.). Vol.~\bibinfo{volume}{9112}.
  \bibinfo{publisher}{Springer International Publishing},
  \bibinfo{address}{Cham}, \bibinfo{pages}{726--729}.
\newblock
\showISBNx{978-3-319-19772-2 978-3-319-19773-9}
\urldef\tempurl%
\url{http://link.springer.com/10.1007/978-3-319-19773-9_102}
\showURL{%
\tempurl}
\newblock
\shownote{Series Title: Lecture Notes in Computer Science.}


\bibitem[\protect\citeauthoryear{on~Computing~Curricula}{on~Computing~Curricula}{2014}]%
        {acm2014curriculum}
\bibfield{author}{\bibinfo{person}{The Joint Task~Force on
  Computing~Curricula}.} \bibinfo{year}{2014}\natexlab{}.
\newblock \bibinfo{booktitle}{\emph{Curriculum Guidelines for Undergraduate
  Degree Programs in Software Engineering}}.
\newblock \bibinfo{type}{{T}echnical {R}eport}. \bibinfo{address}{New York, NY,
  USA}.
\newblock


\bibitem[\protect\citeauthoryear{Parsons and Haden}{Parsons and Haden}{2006}]%
        {parsons2006parson}
\bibfield{author}{\bibinfo{person}{Dale Parsons} {and}
  \bibinfo{person}{Patricia Haden}.} \bibinfo{year}{2006}\natexlab{}.
\newblock \showarticletitle{Parson's Programming Puzzles: A Fun and Effective
  Learning Tool for First Programming Courses}. In
  \bibinfo{booktitle}{\emph{Proceedings of the 8th Australasian Conference on
  Computing Education - Volume 52}} (Hobart, Australia)
  \emph{(\bibinfo{series}{ACE '06})}. \bibinfo{publisher}{Australian Computer
  Society, Inc.}, \bibinfo{address}{AUS}, \bibinfo{pages}{157–163}.
\newblock
\showISBNx{1920682341}


\bibitem[\protect\citeauthoryear{Poulsen, Kulkarni, Herman, and West}{Poulsen
  et~al\mbox{.}}{2022}]%
        {poulsen2022efficient}
\bibfield{author}{\bibinfo{person}{Seth Poulsen}, \bibinfo{person}{Shubhang
  Kulkarni}, \bibinfo{person}{Geoffrey Herman}, {and} \bibinfo{person}{Matthew
  West}.} \bibinfo{year}{2022}\natexlab{}.
\newblock \bibinfo{title}{Efficient Partial Credit Grading of Proof Blocks
  Problems}.
\newblock
\newblock
\urldef\tempurl%
\url{https://doi.org/10.48550/ARXIV.2204.04196}
\showDOI{\tempurl}


\bibitem[\protect\citeauthoryear{Poulsen, Viswanathan, Herman, and
  West}{Poulsen et~al\mbox{.}}{2021}]%
        {poulsen2021evaluating}
\bibfield{author}{\bibinfo{person}{Seth Poulsen}, \bibinfo{person}{Mahesh
  Viswanathan}, \bibinfo{person}{Geoffrey~L. Herman}, {and}
  \bibinfo{person}{Matthew West}.} \bibinfo{year}{2021}\natexlab{}.
\newblock \showarticletitle{Evaluating Proof Blocks Problems as Exam
  Questions}. In \bibinfo{booktitle}{\emph{Proceedings of the 2021 ACM
  Conference on International Computing Education Research}}.
\newblock


\bibitem[\protect\citeauthoryear{Severance, Hanss, and Hardin}{Severance
  et~al\mbox{.}}{2010}]%
        {severance2010ims}
\bibfield{author}{\bibinfo{person}{Charles Severance}, \bibinfo{person}{Ted
  Hanss}, {and} \bibinfo{person}{Josepth Hardin}.}
  \bibinfo{year}{2010}\natexlab{}.
\newblock \showarticletitle{Ims learning tools interoperability: Enabling a
  mash-up approach to teaching and learning tools}.
\newblock \bibinfo{journal}{\emph{Technology, Instruction, Cognition and
  Learning}} \bibinfo{volume}{7}, \bibinfo{number}{3-4} (\bibinfo{year}{2010}),
  \bibinfo{pages}{245--262}.
\newblock


\bibitem[\protect\citeauthoryear{Team}{Team}{2021a}]%
        {orderblocksdocs}
\bibfield{author}{\bibinfo{person}{PrairieLearn Team}.}
  \bibinfo{year}{2021}\natexlab{a}.
\newblock \bibinfo{title}{pl-order-blocks Documentation}.
\newblock
\newblock
\urldef\tempurl%
\url{https://prairielearn.readthedocs.io/en/latest/elements/#pl-order-blocks-element}
\showURL{%
\tempurl}


\bibitem[\protect\citeauthoryear{Team}{Team}{2021b}]%
        {pldocs}
\bibfield{author}{\bibinfo{person}{PrairieLearn Team}.}
  \bibinfo{year}{2021}\natexlab{b}.
\newblock \bibinfo{title}{PrairieLearn Documentation}.
\newblock
\newblock
\urldef\tempurl%
\url{https://prairielearn.readthedocs.io/en/latest/}
\showURL{%
\tempurl}


\bibitem[\protect\citeauthoryear{The Joint Task Force~on Computing~Curricula
  and Society}{The Joint Task Force~on Computing~Curricula and Society}{2016}]%
        {acm2016curriculum}
\bibfield{author}{\bibinfo{person}{Association for Computing Machinery~(ACM)
  The Joint Task Force~on Computing~Curricula} {and}
  \bibinfo{person}{IEEE~Computer Society}.} \bibinfo{year}{2016}\natexlab{}.
\newblock \bibinfo{booktitle}{\emph{Curriculum Guidelines for Undergraduate
  Degree Programs in Computer Engineering}}.
\newblock \bibinfo{type}{{T}echnical {R}eport}. \bibinfo{address}{New York, NY,
  USA}.
\newblock


\bibitem[\protect\citeauthoryear{Weinman, Fox, and Hearst}{Weinman
  et~al\mbox{.}}{2020}]%
        {weinman2020exploring}
\bibfield{author}{\bibinfo{person}{Nathaniel Weinman}, \bibinfo{person}{Armando
  Fox}, {and} \bibinfo{person}{Marti Hearst}.} \bibinfo{year}{2020}\natexlab{}.
\newblock \showarticletitle{Exploring challenging variations of parsons
  problems}. In \bibinfo{booktitle}{\emph{Proceedings of the 51st ACM Technical
  Symposium on Computer Science Education}}. \bibinfo{pages}{1349--1349}.
\newblock


\bibitem[\protect\citeauthoryear{Weintrop and Wilensky}{Weintrop and
  Wilensky}{2015}]%
        {weintrop2015block}
\bibfield{author}{\bibinfo{person}{David Weintrop} {and} \bibinfo{person}{Uri
  Wilensky}.} \bibinfo{year}{2015}\natexlab{}.
\newblock \showarticletitle{To block or not to block, that is the question:
  students' perceptions of blocks-based programming}. In
  \bibinfo{booktitle}{\emph{Proceedings of the 14th international conference on
  interaction design and children}}. \bibinfo{pages}{199--208}.
\newblock


\end{thebibliography}
\balance 

\end{document}